\documentclass{ptephy_v1}

\preprintnumber{XXXX-XXXX} 

\usepackage{bm, braket,amsmath,mathtools,comment}
\usepackage{graphicx,color}

\begin{document}

\title{Transition between vacuum and finite-density states in the infinite-dimensional Bose-Hubbard model with spatially inhomogeneous dissipation }

\author{Shiono Asai$^{1,2,*}$}
\author{Shimpei Goto$^{3,\dag}$}
\author{Ippei Danshita$^{1,\ddag}$
}

\affil{$^1$Department of Physics, Kindai University, Higashi-Osaka, Osaka 577-8502, Japan
\\
$^2$Department of Physics, Nara Women's University, Nara, Nara 630-8506, Japan
\\
$^3$College of Liberal Arts and Sciences, Tokyo Medical and Dental University, Ichikawa, Chiba 272-0827, Japan
\\
{$^*$vas\_asai@cc.nara-wu.ac.jp}
\\
{$^\dag$goto.las@tmd.ac.jp}
\\
{$^\ddag$danshita@phys.kindai.ac.jp}
}
\date{\today}

\begin{abstract}%
We analyze dynamics of the infinite-dimensional Bose-Hubbard model with spatially inhomogeneous dissipation in the hardcore boson limit by solving the Lindblad master equation with use of the Gutzwiller variational method. We consider dissipation processes that correspond to inelastic light scattering in the case of Bose gases in optical lattices. We assume that the dissipation is applied to a half of lattice sites in a spatially alternating manner. We focus on steady states at which the system arrives after long-time evolution. We find that when the average particle density is varied, the steady state exhibits a transition between a state in which the sites without dissipation are vacuum and that containing a finite number of particles at those sites. 
We associate the transition with the tendency of the sites with dissipation towards a local state at infinite temperature.
\end{abstract}

\subjectindex{A63, I22}

\maketitle

\section{Introduction
\label{sec:introduction}}
Recent technological advances in cold-atom experiments have provided unique opportunities for analyzing open quantum many-body systems with widely controllable dissipation and stimulated theoretical studies on this topic~\cite{daley,sieberer,ashida}. Various kinds of controllable dissipation have been realized thus far, such as one-body losses~\cite{barontini,takasu,ren}, two-body losses~\cite{tomita}, and inelastic light scattering~\cite{patil,luschen,bouganne}.

Previous theoretical studies~\cite{diehl-1,tomadin} have considered dynamics of a Bose-Hubbard system with dissipation processes which tend to lock the phase difference of the particle field between nearest neighboring sites. It has been shown that steady states reached after long-time evolution exhibit a transition between a Bose condensed state and a non-condensed state when the dissipation strength is varied. Such transition phenomena in Bose-Hubbard systems with dissipation have recently attracted renewed interest in connection with the transitions of the scaling law of entanglement induced by probabilistic measurements or dissipation~\cite{tang,goto,fuji,regemortel}. It is natural to ask whether similar transition phenomena can occur for other types of dissipation.

In this paper, we consider dissipation processes corresponding to the inelastic light scattering in experiments with ultracold gases~\cite{patil,luschen,bouganne}, in order to explore the transitions of the steady states in a Bose-Hubbard system at infinite spatial dimensions. The infinite dimensional system is suited for such exploration in the sense that its exact quantum dynamics can be analyzed within the Gutzwiller-type mean-field treatment.  Specifically, we consider dissipation of a spatially alternating pattern, in which all the sites with dissipation are neighboring to those without dissipation and vice versa. By solving the Lindblad master equation with use of the Gutzwiller variational method, we analyze long-time dynamics of the dissipative Bose-Hubbard model at infinite dimensions in the hardcore boson limit. Taking a superfluid ground state, which is spatially homogeneous, as an initial state, we calculate the time evolution of the local particle density.
We find that there occurs in the steady states a transition when the average particle density is varied.
We discuss a mechanism of how the transition occurs, focusing on the absorption of particles into lattice sites with the dissipation.

The rest of the paper is organized as follows.
In Sec.~\ref{sec:model},  we explain the Bose-Hubbard model, which describes ultracold bosons in a deep optical lattice,  and the master equation of the Lindblad form.
In Sec.~\ref{sec:method}, we review the Gutzwiller variational method used for analyzing Lindblad dynamics of the dissipative Bose-Hubbard model. 
In Sec.~\ref{sec:result}, we analyze the dissipative dynamics starting from a superfluid ground state and discuss transitions of the steady states reached after long-time evolution.
In Sec.~\ref{sec:summaries}, we summarize the results.

\section{Model}
\label{sec:model}
We consider the $D$-dimensional Bose-Hubbard model \cite{fisher,jaksch} on a hypercubic lattice,
\begin{equation}
\hat{H} = - J \sum_{\langle j,l\rangle} ( \hat{b}_j^{\dag} \hat{b}_l+ \hat{b}_l^{\dag}  \hat{b}_j)+\frac{U}{2}\sum_{j}\hat{n}_j ( \hat{n}_j - 1) - \mu \sum_{j} \hat{n}_j.
\end{equation}
Here, $\hat{b}_j$ ($\hat{b}^{\dag}_j$) is the annihilation (creation) operator of a boson at site $j$, $\hat{n}_j=\hat{b}^{\dag}_j\hat{b}_j$, and $\langle j,l \rangle$ denotes nearest-neighboring pairs of the sites.
$J$ is the hopping parameter, $U$ is the interaction strength, and $\mu$ is chemical potential.
The ground state of this model can be either superfluid or Mott-insulator phase~\cite{fisher}. 
If the number of particles per site $\nu$ is an integer and the dimensionless quantity $zJ/U$ is smaller than a certain critical value, the ground state is the Mott insulator phase, where $z=2D$ is the coordination number.
Otherwise, it is the superfluid phase. 
In Sec.~\ref{sec:result}, we study dissipative quantum dynamics starting from a superfluid ground state.

We anticipate dissipative dynamics of the density matrix $\hat{\rho}$ of the Bose-Hubbard system
described by the master equation in  Lindblad form~\cite{lindbrad}
\begin{equation}
\label{eq:lindblad}
\frac{d\hat{\rho}}{dt}= - i [\hat{H},\hat{\rho}]+\frac{1}{4}\sum_{j}\Gamma_j( - \hat{L}_j^{\dag}\hat{L}_j\hat{\rho} - \hat{\rho}\hat{L}_j^{\dag}\hat{L}_j+2\hat{L}_j\hat{\rho}\hat{L}_j^{\dag}).
\end{equation}
Here, $\hat{L}_j$ is the Lindblad operator describing dissipation processes that occur at site $j$ and the constant $\Gamma_j$ is the rate of the dissipation. 
In addition, we set $\hbar =1$. In the present study, we assume a simple form of the dissipation, i.e., $\hat{L}_j = \hat{n}_j$. In systems of ultracold atomic gases, this type of dissipation corresponds to inelastic scattering between an atom and a photon~\cite{patil,luschen,bouganne}.

\section{Methods}
\label{sec:method}
\subsection{Gutzwiller variational method}
In order to analyze Lindblad dynamics of the dissipative Bose-Hubbard model, we use the Gutzwiller variational method~\cite{rokhsar,diehl-1}. It is a kind of mean field approximation, in which a  quantum many-body state is expressed as a direct product state of a certain local states.
Owing to its mean-field nature, the accuracy of this approximation improves as the spatial dimension $D$ becomes higher. It is exact in the limit of $D\rightarrow \infty$.
There are several previous studies that have applied the Gutzwiller approximation to the dynamics of the dissipative Bose-Hubbard model~\cite{tomita, diehl-1, tomadin, leboite, vidanovic}.
In particular, it has been used in Refs.~\cite{diehl-1, tomadin} to show that a dynamical phase transition occurs when one increases the strength of dissipation processes that tend to lock the phase difference of the particles between two adjacent sites. Moreover, Vidanovi\'c {\it et al.}~have incorporated local single-particle losses within the Gutzwiller approximation~\cite{vidanovic}. We employ this approximation in order to explore dynamical phase transitions induced by spatially inhomogeneous light scattering.

In the Gutzwiller variational method,
the density matrix $\hat{\rho}$ is 
expressed as
\begin{equation}
\hat{\rho}= \prod_{\otimes j} \sum_{n,m}c_{j,n,m}\ket{n}_j\bra{m}_j
\label{eq:rho}
\end{equation}
The summation with respect to the number of particles at each site ($n$ and $m$) continues until infinity in general. Nevertheless, in many practical situations, including those that we consider in the present study, the occupation probabilities of large-$n$ states are negligibly small so that one can safely assume that the maximum number of particles at each site $n_{\rm max}$ is finite.
Using the density matrix of Eq.~(\ref{eq:rho}), we define the local superfluid order parameter as
 $\Psi_j=\braket{\hat{b}_j}=\mathrm{Tr}[\hat{\rho}\hat{b}_j]$.
In the Gutzwiller variational method, the second-order fluctuations of the field operators from the mean fields are ignored so that the Hamiltonian is approximated as
\begin{equation}
\hat{H}\simeq\sum_{j}\hat{H}_j^{\rm GW}
\label{eq:GWhamil}
\end{equation}
where
\begin{align}
\begin{aligned}
\hat{H}_j^{\rm GW}&=\hat{H}_{0,j}^{\rm loc}+\hat{H}_{{\rm hop},j}^{\rm GW}\\
\hat{H}_{0,j}^{\rm loc}&=\sum_j\frac{U}{2}\hat{n}_j(\hat{n}_j-1) - \mu\sum_j\hat{n}_j\\
\hat{H}_{{\rm hop},j}^{\rm GW}&=-J\sum_{\langle l\rangle_j}(\Psi^*_l\hat{b}_j+\hat{b}_{j}^{\dag}\Psi_l)\\
\end{aligned}
\end{align}
$\sum_{\langle l\rangle_j}$ means the summation with respect to sites neighboring to site $j$.
Substituting Eq.~(\ref{eq:GWhamil}) into Eq.~(\ref{eq:lindblad}), we obtain
\begin{equation}
\frac{d\hat{\rho}_j^{\rm GW}}{dt}= - i[\hat{H}_j^{\rm GW},\hat{\rho}_j^{\rm GW} ]+L^{\rm loc}(\hat{\rho}_j^{\rm GW})
\label{eq:lindbladGW}
\end{equation}
where
\begin{equation}
L^{\rm loc}(\hat{\rho}_j^{\rm GW})=\frac{1}{4}\sum_{j}\Gamma_j( - \hat{L}_j^{\dag}\hat{L}_j\hat{\rho_j} - \hat{\rho_j}\hat{L}_j^{\dag}\hat{L}_j+2\hat{L}_j\hat{\rho_j}\hat{L}_j^{\dag}).
\end{equation}
One of the main advantages of the Gutzwiller approximation is that since the field operators in the hopping term are replaced with their mean values, $\hat{H}_j^{\rm GW}$ consists only of local operators at site $j$. The master equation within the Gutzwiller approximation of Eq.~(\ref{eq:lindbladGW}) is thereby represented in the site-decoupled form. This decoupling allows us to solve the master equation at a very low numerical cost.

In the following calculations, we assume the limit of the infinite dimensions ($D\rightarrow \infty$), in which the Gutzwiller variational method is exact, while keeping $zJ$ to be finite. We use the fourth-order Runge-Kutta method for solving the master equation. We set the units of the time and the energy to be $\hbar/(zJ)$ and $zJ$. In order to show that dynamical transitions can occur in the presence of inhomogeneous light scattering, we assume that the system is in the hardcore boson limit ($U\rightarrow \infty$), i.e., $n_{\rm max} = 1$ and that the initial state for time evolution is the ground state for a given value of $\nu$. We also assume that $0<\nu<1$ so that the initial state is a homogeneous superfluid state. The validity of the hardcore boson approximation for systems with finite $U/(zJ)$ will be discussed in Appendix A. 
\section{Results}
\label{sec:result}
%

\subsection{Homogeneous dissipation}
\label{whole}
%
In this subsection, we consider the case that the dissipation corresponding to inelastic light scattering processes is spatially homogeneous ($\Gamma_j = \Gamma$) in order to numerically confirm that 
there is no transition of the steady state caused by varying $\nu$ in this case.

\begin{figure}
  \centering
  \includegraphics[scale=0.25]{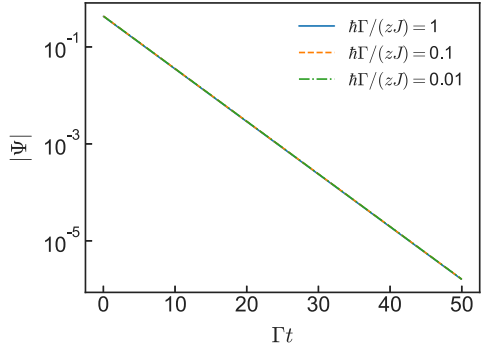}
  \caption{
  Time evolution of the amplitude of the superfluid order parameter $|\Psi|$ for homogeneous dissipation, where $\nu=0.25$, and $ \hbar\Gamma/(zJ)= 0.01,0.1,1$. The vertical axis is shown in a logarithmic scale.
The three curves coincide with one another since $1/\Gamma$ is chosen as the unit of time.
    \label{fig:zentaipsi}}
  \end{figure}

In Fig.~\ref{fig:zentaipsi}, we set $\nu = 0.25$ and $\Gamma = 0.01,0.1,1$, and the time evolution of the amplitude of the superfluid order parameter $\Psi_j = \Psi$ is plotted as a function of $\Gamma t$.
Regardless of the strength of the dissipation, as long as $\Gamma$ is finite, $\Psi$ exponentially decreases to zero over time. 
At the end of the long-time evolution, at which $ \Psi $ is zero, the local density matrix is diagonal in the local Fock basis. Since this happens regardless of the value of $\nu$, no transition occurs with varying $\nu$.

%
%
\subsection{Spatially alternating dissipation}
\label{1/4site}
\begin{figure}[t]
\centering
\includegraphics[scale=0.2]{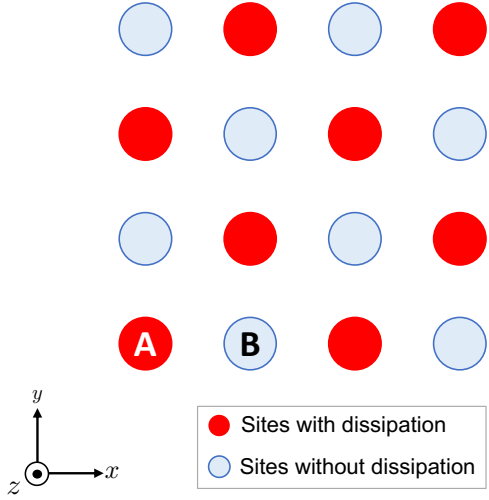}
\caption{Schematic picture of spatially alternating dissipation for $D=2$. The dissipation term $\Gamma_j$ is finite at the red-colored sites while it is zero at the blue-colored sites.  The numbers written on the sites of the unit cell ($\rm{A}$ and $\rm{B}$) represent the sublattice index $j$. The configuration spreads homogeneously in the $z$-axis direction. 
\label{fig:illustration}}
\end{figure}

%
%
In this subsection, we show a simple example in which 
there is a transition of the steady state caused by varying $\nu$ when the dissipation is applied partly to the system.
Specifically, we consider spatially alternating dissipation, in which there is a two-sublattice structure. We assume that all the sites neighboring to sublattice $\rm{A}$ ($\rm{B}$) belong to sublattice $\rm{B}$ ($\rm{A}$). For illustration, the spatial configuration of the sites with dissipation for $D=2$ is shown in Fig.~\ref{fig:illustration}.  The dissipation with strength $\Gamma_j = \Gamma$ is imposed to sublattice $\rm{A}$ while $\Gamma_j = 0$ at the other sublattice.

\begin{figure}[tb]
  \centering
     \includegraphics[scale=0.4]{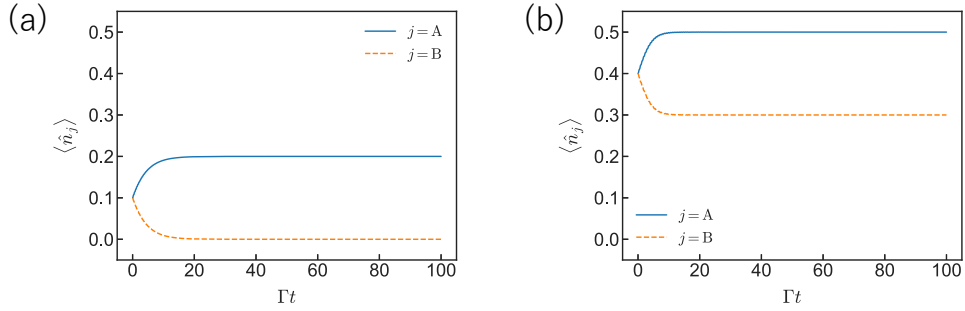}
   \caption{
Time evolution of the particle density $\langle \hat{n}_{j} \rangle $ for (a) $\nu=0.1$ and (b) $\nu=0.4$, where $\hbar\Gamma/(zJ)=0.1$. 
The solid blue and dashed orange lines correspond to $j=\rm{A}$ and $j=\rm{B}$ respectively. 
\label{fig:dens}}
\end{figure}

Figure~\ref{fig:dens} shows the time evolution of the local particle density $\langle \hat{n}_j \rangle$
for $\nu=0.1$ and $0.4$ at $\Gamma=0.1$. 
As shown in Fig.~\ref{fig:dens}(a), $\langle \hat{n}_j \rangle$ at the sublattice without dissipation ($j=\rm{B}$) decreases to zero over time when $\nu = 0.1$.
In contrast, we see in Fig.~\ref{fig:dens}(b), where $\nu = 0.4$, $\langle \hat{n}_j \rangle$ at $j=\rm{B}$ remains finite even after the long-time evolution. In both cases, $\langle \hat{n}_j \rangle$ at the sublattice with dissipation ($j=\rm{A}$) monotonically increases to converge to a certain value.

\begin{figure}[t]
\centering
\includegraphics[scale=0.3]{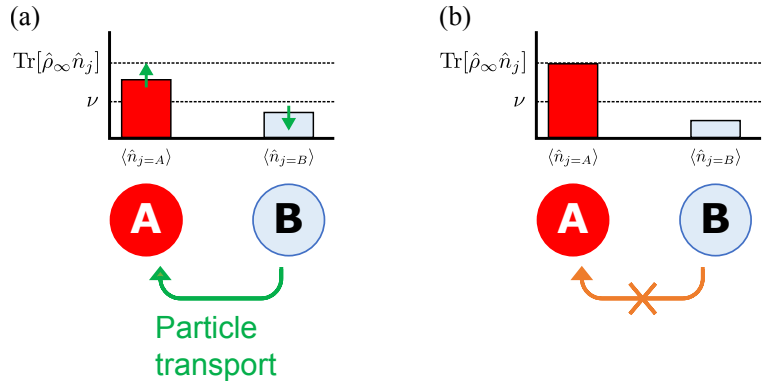}
\caption{Schematic illustration of the particle transport within the unit cell during the dynamics in the presence of local dissipation.
(a) When the initial particle number per site $\nu$ is sufficiently small, the sublattice with dissipation ($j=\rm{A}$) absorbs the particles from the other sublattices until the latter sublattices become empty.
(b) When $\nu$ is too large, the local density matrix at sublattice $j=\rm{A}$ reaches a state with infinite temperature before the other sublattices become empty. The particle transport stops afterward.
\label{fig:particlemovement}}
\end{figure}

Figure~\ref{fig:particlemovement} illustrates the reason why the above-mentioned behavior of $\langle \hat{n}_j\rangle$ appears.
As shown in Fig.~\ref{fig:dens},  the dissipation causes the intersite transport of particles.
In Fig.~\ref{fig:dens}(a), where $\nu = 0.1$, while $\braket {\hat {n}_{j}}>\nu$ on the sublattice with dissipation ($ j= \rm{A} $), $ \braket {\hat{n}_{j}}=0$ on the sublattices  without dissipation ($ j= \rm{B}$).
As illustrated in Fig.~\ref{fig:particlemovement}(a), the sites with dissipation absorb the particles from the other sites, because the local density matrices at the former sites tend to approach a perfectly mixed state with $\langle n_{j=\rm{A}} \rangle = 0.5$, which is larger than the initial density.
The fact that the sites without dissipation become empty ($\braket {\hat {n}_{j}} = 0 $ for $j=\rm{B}$) means that  the particles are exhausted before the local density matrices at $j=\rm{A}$ reach the state with infinite temperature. 
In this way, all the particles at the sites without dissipation that had been present at the beginning of the dissipative dynamics are absorbed to the sites with dissipation ($j=\rm{A}$).

When $\nu$ is relatively large as shown in  Fig.~\ref{fig:dens}(b),  
 $ \braket {\hat {n} _ {j}} = 0.5 $  on the sites with dissipation after the long-time evolution.
This means that the local density matrix is the state at infinite temperature and that the associated saturation of the particle number at $j=\rm{A}$ ceases the particle transport.  
As a consequence, there remain the particles at the sites without dissipation.

\begin{figure}[t]
\centering
\includegraphics[scale=0.4]{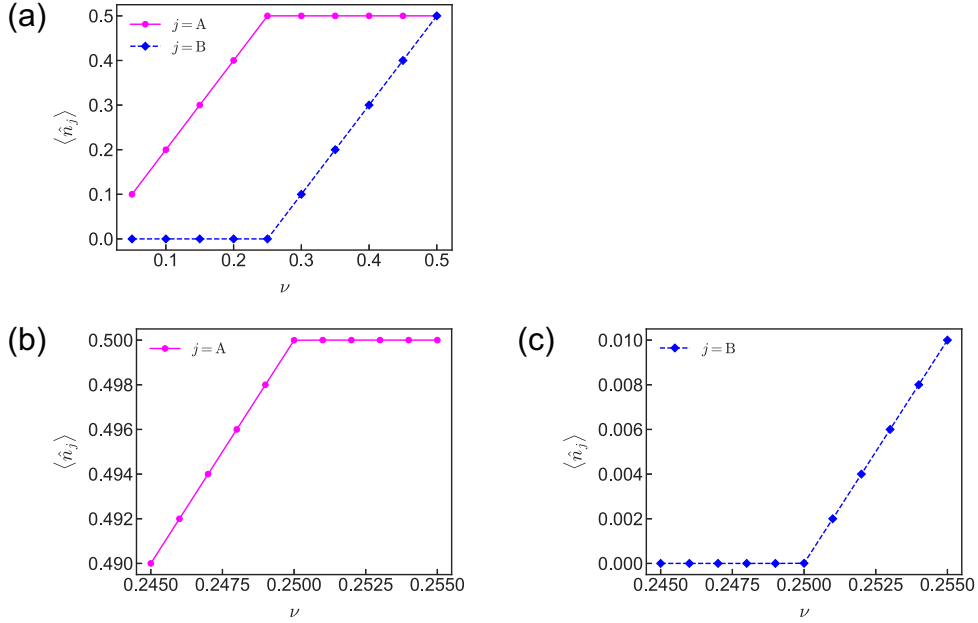}
\caption{(a) The squared amplitude of the 
local particle density $\langle \hat{n}_j\rangle$ after the long-time evolution versus the number of particles per site $\nu$ for $j=\rm{A}$ and $\rm{B}$. 
(b) and (c) are magnified views of (a) near the transition point for $j=\rm{A}$ and $\rm{B}$, respectively.
\label{fig:souzu}}
\end{figure}

The difference between the two cases, namely $\nu = 0.1$ and $0.4$, raises an important question of whether the change from a state in which the sites without dissipation are empty to a state in which a finite number of particles occupy those sites occurs as a transition or a crossover when $\nu$ is increased.
In order to answer this question, we plot 
$\langle \hat{n}_j\rangle$ after the long-time evolution
as functions of $\nu$ in Fig.~\ref{fig:souzu}.
Notice that the results shown in Fig.~\ref{fig:souzu} are independent of the dissipation strength $\Gamma$.
We see in Fig.~\ref{fig:souzu} that the sites without dissipation are empty, i.e., they are in a vacuum state at $\nu <0.25$, because the sites with dissipation can accommodate all the particles.
In contrast, they contain a finite number of particles at $\nu >0.25$.
As clearly shown in Figs.~\ref{fig:souzu}(b)  and (c), the first derivative of 
$\langle \hat{n}_j\rangle$ with respect to $\nu$ is discontinuous ($\langle \hat{n}_j\rangle$ forms a kink) at $\nu = 0.25$, implying that the change at $\nu = 0.25$ is a transition that is similar to a continuous one in the classification of standard phase transitions in static systems.

\section{Summaries\label{sec:summaries}}
In conclusion, we have investigated non-equilibrium dynamics of the infinite-dimensional Bose-Hubbard model with dissipation processes corresponding to inelastic light scattering, focusing on effects of spatial inhomogeneity of the dissipation. Specifically, we considered dissipation of a spatially alternating pattern with two-sublattice structure.
We solved the Lindblad master equation for the dissipative Bose-Hubbard model in the hardcore boson limit within the Gutzwiller approximation to calculate the time evolution of 
the local particle density. 
We found that the steady state reached after long-time evolution exhibits a transition
when the average particle density of the initial state increases. 
Associated with the transition, the steady state changes from a state in which the sites without dissipation are empty to that in which the local particle density at the sites without dissipation is finite.

While throughout the paper we restricted ourselves to dynamics of the Bose-Hubbard model at infinite dimensions described by the Gutzwiller variational method, in which entanglement in the real space is irrelevant, it will be interesting to take into account effects of the entanglement in the future study. One possible approach is an extension of the Gutzwiller approximation using the cluster mean-field theory~\cite{yamamoto}, which includes the entanglement within the cluster. Quasi-exact numerical simulations of one-dimensional systems by means of matrix-product states~\cite{daley,goto} may be another candidate approach.

\section*{Acknowledgment}
We thank M.~Kunimi and D.~Yamamoto for useful discussions. This work was financially supported by JST CREST (Grant No.\ JPMJCR1673), by JST FOREST (Grant No.\ JPMJFR202T), by MEXT Q-LEAP Grant No.\ JPMXS0118069021, KAKENHI from Japan Society for Promotion of Science: Grant No.\ JP18H05228, No.\ JP20K14377, JP21H01014, and JP21H05185.
\vspace{10mm}
\appendix
\counterwithin{figure}{section}
\hspace{-3mm}{\bf Appendix:}
\section{Simulation without the hardcore constraint}
\label{sec:supplement}
\begin{figure}
  \centering
   \includegraphics[scale=0.25]{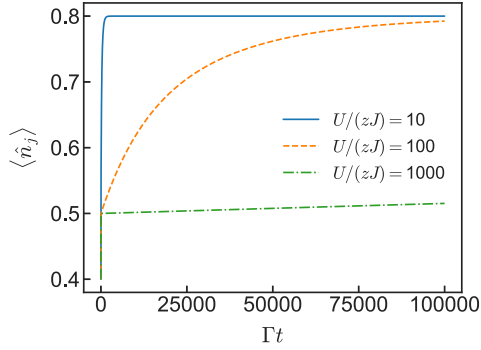}
  \caption{Time evolution of the particle density $\langle \hat{n}_{j} \rangle$ at site $j= \rm{A}$ in the range up to $zJt/\hbar = 1\times 10^5$, where $\nu=0.4$, $ \hbar\Gamma/(zJ) =0. 1 $, and $ n_{\mathrm{max}} =2$.
The solid blue, dashed orange, and dashed-dotted green correspond to $U/(zJ)=10,100$, and $1000$, respectively.
  \label{fig:softcore}}
\end{figure}

\begin{figure}[t]
   \centering
   \includegraphics[scale=0.25]{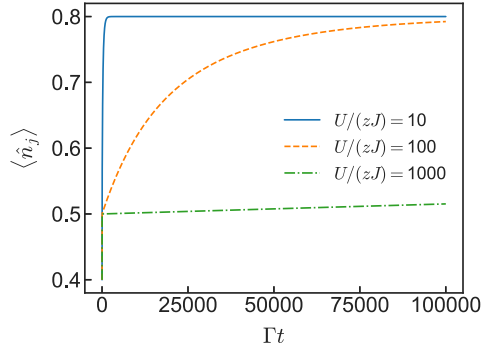}
  \caption{
 Time evolution of the particle density $\langle \hat{n}_{j} \rangle$ in the range up to $zJt/\hbar= 1000$ where $ \nu = 0.4 $, $ n_{\mathrm {max}} = 2 $,  and $ \hbar\Gamma/(zJ) = 0.1 $. The solid blue, dashed orange, dashed-dotted green, and dotted red lines correspond to $U/(zJ)=10, 20,50$, and $100$, respectively.
  \label{fig:softcorelarge}}
 \end{figure}

We discuss the validity of the hardcore boson approximation for systems with finite $U$. Here we take $zJ$ and $\hbar/(zJ)$ as units of the energy and the time.
We consider the situation discussed in Sec.~\ref{1/4site} in which the dissipation is imposed in a spatially alternating manner.
 We set $ n_{\mathrm{max}}=2$ and change the interaction from $U=10$ to $1000$.
Figures~\ref {fig:softcore} and \ref{fig:softcorelarge} represent the time evolution of the average number of particles $\langle \hat{n}_{j}\rangle$ at site $j=\rm{A}$, where $\Gamma=0.1$ and  $\nu=0.4$.
In the case that $U=1000$, after $\langle \hat{n}_{j=\rm{A}}\rangle$ increases from 0.4 to 0.5, it remains almost constant up to the very long time, namely $\Gamma t =1\times 10^5$. This means that the local density matrix at $j=\rm{A}$ stays in the completely mixed state for the hardcore boson ($n_{\rm max} = 1$) in this very long timescale.   
When $U$ decreases, the deviation of $\langle \hat{n}_{j=\rm{A}}\rangle$ from 0.5 gradually increases. 
In the case that $U=10$ shown in Fig.~\ref{fig:softcore},  $\langle \hat{n}_{j=\rm{A}}\rangle$ converges to $0.8$, which is the value of $\langle \hat{n}_{j=\rm{A}}\rangle$ in the steady state for $n_{\rm max} = 2$ and $\nu =0.4$.

%
%
 \begin{figure}[t]
   \centering
   \includegraphics[scale=0.25]{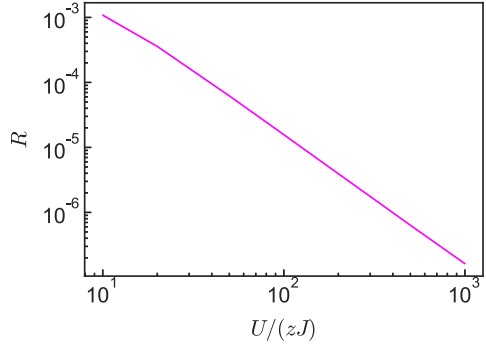}
  \caption{The slope $R$ of the linear growth of $\langle \hat{n}_{j=\rm{A}}\rangle$ as a function of $U/(zJ)$. $R$ is extracted from the numerical data of $\langle \hat{n}_{j=\rm{A}}\rangle$ in the time domain where $\langle \hat{n}_{j=\rm{A}}\rangle$ increases from 0.5 to 0.8.
  \label{fig:softcorekatamuki}}
 \end{figure}

$\langle \hat{n}_{j=\rm{A}}\rangle$ grows almost linearly with time after it reaches 0.5. In order to quantify the deviation from the hardcore boson approximation, we extract the slope of the linear growth $R$ of $\langle \hat{n}_{j=\rm{A}}\rangle$ from the numerical data.  
Figure~\ref{fig:softcorekatamuki} shows the slope $R$ as a function of $U$. We find that $R$ decreases exponentially with $U$, meaning that the timescale for the hardcore boson approximation to be valid increases exponentially with increasing $U$.

\newpage
\bibliographystyle{apsrev4-2}

\end{document}